\documentclass[%
superscriptaddress,
 amsmath,amssymb,
 superscriptaddress,
 aps,
 prl,
 twocolumn
]{revtex4-2}
\usepackage{graphicx}%
\usepackage{dcolumn}
\usepackage{bm}
\usepackage{float}
\usepackage{xcolor}

\begin{document}

\title{Universal mechanical response of metallic glasses during strain-rate-dependent uniaxial compression}
\author{Weiwei Jin}
\affiliation{Department of Mechanical Engineering and Materials Science, Yale University, New Haven, Connecticut 06520, USA}
\author{Amit Datye}
\affiliation{Department of Mechanical Engineering and Materials Science, Yale University, New Haven, Connecticut 06520, USA}
\author{Udo D.\ Schwarz}
\affiliation{Department of Mechanical Engineering and Materials Science, Yale University, New Haven, Connecticut 06520, USA}
\affiliation{Department of Chemical and Environmental Engineering, Yale University, New Haven, Connecticut 06520, USA}
\author{Mark D.\ Shattuck}
\affiliation{Benjamin Levich Institute and Physics Department, The City College of New York, New York, New York 10031, USA}
\author{Corey S.\ O'Hern}
 \email{corey.ohern@yale.edu}
\affiliation{Department of Mechanical Engineering and Materials Science, Yale University, New Haven, Connecticut 06520, USA}
\affiliation{Department of Physics, Yale University, New Haven, Connecticut 06520, USA}
\affiliation{Department of Applied Physics, Yale University, New Haven, Connecticut 06520, USA}
\affiliation{Graduate Program in Computational Biology and Bioinformatics, Yale University, New Haven, Connecticut 06520, USA}

\begin{abstract}

Experimental data on the compressive strength $\sigma_{\rm max}$ versus strain rate ${\dot \varepsilon}_{\rm eng}$ for metallic glasses undergoing uniaxial compression shows significantly different behavior for different alloys. For some metallic glasses, $\sigma_{\rm max}$ decreases with increasing ${\dot \varepsilon}_{\rm eng}$, for others, $\sigma_{\rm max}$ increases with increasing ${\dot \varepsilon}_{\rm eng}$, and for others $\sigma_{\rm max}$ versus ${\dot \varepsilon}_{\rm eng}$ is nonmonotonic. Using numerical simulations of metallic glasses undergoing uniaxial compression at nonzero strain rate and temperature, we show that they obey a universal relation for the compressive strength versus temperature, which determines their mechanical response.  At low ${\dot \varepsilon}_{\rm eng}$, increasing strain rate leads to increases in temperature and {\it decreases} in $\sigma^*_{\rm max}$, whereas at high ${\dot \varepsilon}_{\rm eng}$, increasing strain rate leads to decreases in temperature and {\it increases} in $\sigma^*_{\rm max}$.  This non-monotonic behavior of $\sigma^*_{\rm max}$ versus temperature causes the nonmonotonic behavior of $\sigma^*_{\rm max}$ versus ${\dot \varepsilon}_{\rm eng}$. Variations in the internal dissipation change the characteristic strain rate at which the nonmonotonic behavior occurs. These results are general for a wide range of metallic glasses with different atomic interactions, damping coefficients, and chemical compositions. 
\end{abstract}
\maketitle

The combination of superior strength and hardness, large elastic limit, and high fracture toughness make bulk metallic glasses (BMGs) a promising materials class for numerous structural applications~\cite{schroers2004,greer2007,demetriou2011,suryanarayana2017}. In contrast to conventional alloys, BMGs are amorphous (i.e. they lack long-range crystalline order), and their response to deformation is not governed by the generation and motion of topological defects. Instead, researchers have shown that shear transformation zones, where atoms undergo collective, non-affine motion, control the mechanical response of metallic glasses. Numerous studies have probed the unique mechanical response of metallic glasses subjected to quasistatic deformations at low temperature. However, understanding the dynamic mechanical response of metallic glasses at finite strain rates and temperatures near the glass transition temperature is important for many engineering applications.

\begin{figure}[htbp]
\centering
\includegraphics[width=0.45\textwidth]{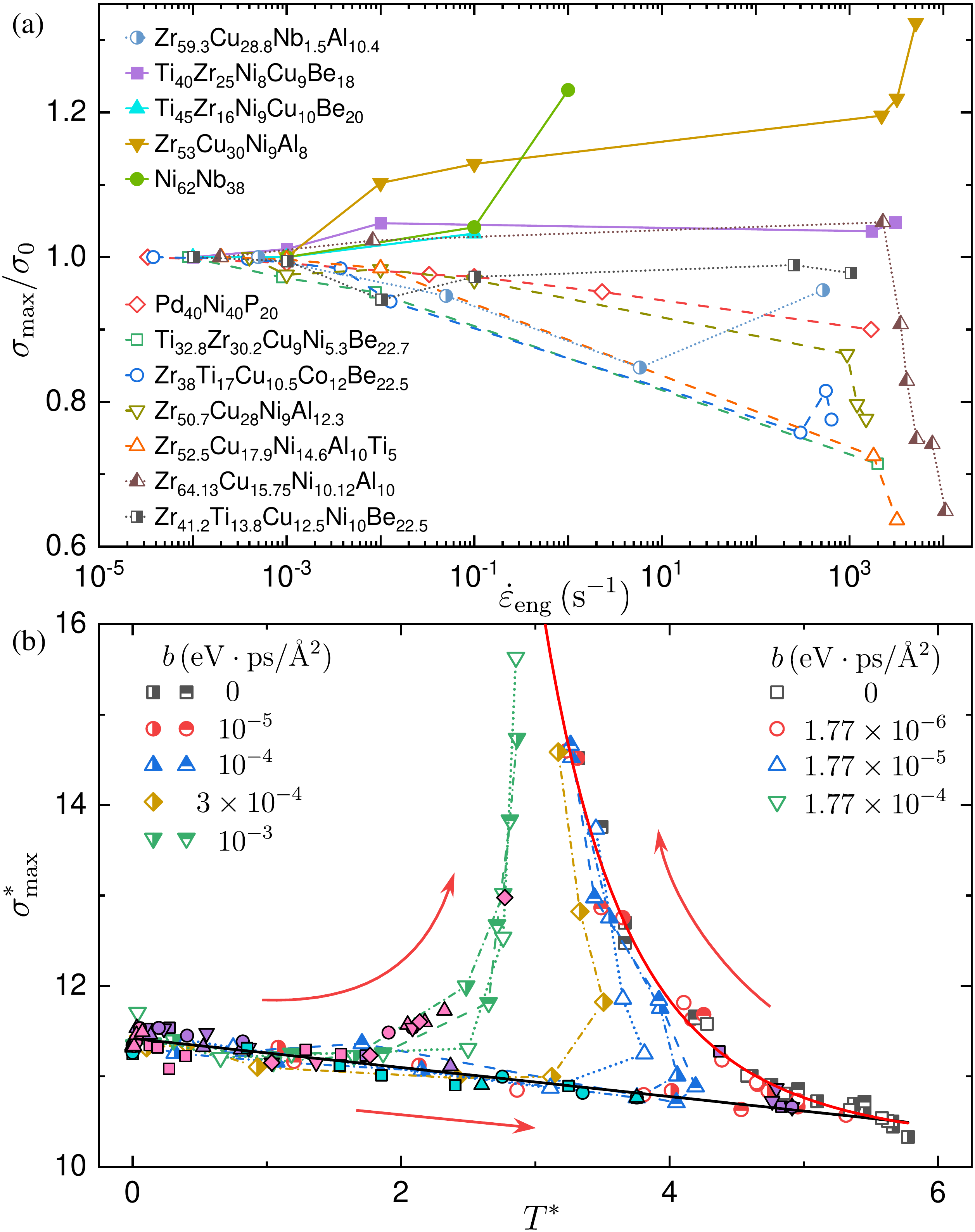}
\caption{(a) Compressive strength $\sigma_{\rm max}$ (normalized by the value $\sigma_0$ at the smallest strain rate) plotted as a function of engineering strain rate ${\dot \varepsilon}_{\rm eng}$ (in $s^{-1}$) from experimental studies of $12$ bulk metallic glasses undergoing uniaxial compression~\cite{zhang2007,ma2009,chen2015,amit2022,mukai2002,xue2008,zheng2011,li2017,wang2014,lu2003,liu2010,ramachandramoorthy2021}. (b) Scaled compressive strength $\sigma^*_{\rm max}$ versus scaled average temperature $T^*$ during uniaxial compression. The EAM simulation results for $\rm Ni_{62}Nb_{38}$ and $\rm Zr_{60}Cu_{29}Al_{11}$ and LJ simulation results for  $\rm Ni_{62}Nb_{38}$ are represented by symbols that are half-filled on the right, half-filled on the top, and open symbols, respectively. Filled symbols with black borders correspond to the experimental results in (a). The arrows indicate the direction of increasing strain rate. The solid lines give the asymptotic behavior in the limit of small damping coefficient. }
\label{figExpSim}
\end{figure}

Uniaxial compression of bulk metallic glass pillars is a common mechanical test that probes their nano- and micro-scale mechanical response.  Studies have shown that the strength of BMGs under compression decreases with increasing temperature~\cite{lu2003,dubach2007,ma2015}, since higher temperatures enhance the activation of shear transformation zones and formation of shear bands~\cite{lewandowski2006}. However, there is no consensus about the behavior of the compressive strength as a function of strain rate for BMGs~\cite{zhang2007,ma2009,chen2015,amit2022,mukai2002,wang2014,xue2008,zheng2011,li2017,lu2003,liu2010,ramachandramoorthy2021}, where the compressive strength is defined as the maximum engineering stress prior to steady flow. For example, as shown in Fig.~\ref{figExpSim} (a), the compressive strength of millimeter-sized Ti- and Zr-based BMGs (such as $\rm Ti_{45}Zr_{16}Ni_{9}Cu_{10}Be_{20}$ \cite{zhang2007}, $\rm Ti_{40}Zr_{25}Ni_{8}Cu_{9}Be_{18}$ \cite{ma2009}, and $\rm Zr_{53}Cu_{30}Ni_{9}Al_{8}$ \cite{chen2015}), and $\rm Ni_{62}Nb_{38}$ \cite{amit2022}, increases with strain rate.  In contrast, the compressive strength of similar Ti- and Zr-based BMGs (e.g. $\rm Ti_{32.8}Zr_{30.2}Cu_{9}Ni_{5.3}Be_{22.7}$ \cite{wang2014}, $\rm Zr_{38}Ti_{17}Cu_{10.5}Co_{12}Be_{22.5}$ \cite{xue2008}, $\rm Zr_{50.7}Cu_{28}Ni_{9}Al_{12.3}$ \cite{zheng2011}, and $\rm Zr_{52.5}Cu_{17.9}Ni_{14.6}Al_{10}Ti_{5}$ \cite{li2017}) and $\rm Pd_{40}Ni_{40}P_{20}$ \cite{mukai2002} decreases with strain rate. In addition, recent studies have shown that the compressive strength for $\rm Zr_{59.3}Cu_{28.8}Nb_{1.5}Al_{10.4}$ is non-monotonic with strain rate; the strength first decreases and then increases with increasing strain rate \cite{ramachandramoorthy2021}. Does the fact that different BMGs possess different strain-rate-dependent compressive strength mean that the mechanical response of these materials depends sensitively on each particular alloy and composition? We seek to identify the dominant mechanism that controls the strain-rate dependent compressive strength so that we can potentially describe the mechanical response of all BMGs to uniaxial compression.

In this Letter, we perform molecular dynamics simulations of metallic glasses undergoing uniaxial compression using the embedded atom method (EAM) and coarse-grained Lennard-Jones (LJ) models to understand their mechanical response.  We find that local temperature fluctuations and internal dissipation mechanisms that occur during uniaxial compression control the behavior of the compressive strength versus strain rate in BMGs.  We show that the compressive strength $\sigma_{\rm max}$ versus the engineering strain rate $\dot \varepsilon_{\rm eng}$ from the experiments in Fig.~\ref{figExpSim} (a) can be mapped onto the data from the computer simulations of metallic glasses (both EAM and LJ) by quantifying the internal damping coefficient and temperature for each BMG. In Fig.~\ref{figExpSim} (b),
we plot the scaled compressive strength $\sigma^*_{\rm max}$ of the BMGs in (a) versus the scaled average temperature $T^*$, where the arrows indicate increasing strain rate. We find universal behavior for the compressive strength for both the experimentally characterized BMGs and computer simulations of BMGs.  In the limit of low damping coefficient ($b\rightarrow 0$), first the compressive strength decreases with increasing strain rate, since increases in strain rate cause increases in temperature and the compressive strength scales as $\sigma^*_{\rm max} = \sigma_1  -\alpha T^*$, where $\sigma_1$ is the compressive strength in the $T \rightarrow 0$ limit and $\alpha \approx 0.16$. When the rescaled temperature reaches $T^* \approx 5.8$, further increases in strain rate cause the rescaled temperature to {\it decrease} and $\sigma^*_{\rm max}$ increases as a power-law with decreasing temperature, $\sigma^*_{\rm max} \sim (T^*)^{-\beta}$, with $\beta \approx 4.8$. For finite damping coefficients, the strain rate-driven temperature dependence of $\sigma^*_{\rm max}$ deviates from the chevron-shaped asymptotic behavior.  In particular, in the limit of high damping coefficient, initial increases in strain rate cause only small increases in temperature, and the compressive strength does not decrease significantly.  With further increases in strain rate, the temperature begins to decrease and the compressive strength increases rapidly, approaching the $\sigma^*_{\rm max} \sim (T^*)^{-\beta}$ asymptote. This work provides a universal description of the wide range of behaviors of the compressive strength versus strain rate found in BMGs undergoing uniaxial compression.  

{\em Methods.}---For the molecular dynamics simulations of uniaxial compression of BMGs,
we focus on two particular BMGs, $\rm Ni_{62}Nb_{38}$ and $\rm Zr_{60}Cu_{29}Al_{11}$ \cite{yokoyama2008}. (Note that experimental measurements of the compressive strength for $\rm Ni_{62}Nb_{38}$ and the related alloy $\rm Zr_{59.3} Cu_{28.8} Nb_{1.5} Al_{10.4}$ are shown in Fig.~\ref{figExpSim} (a).)  For the atomic interactions, we consider EAM potentials, which include many-body interactions from the electronic degrees of freedom and have been developed separately for $\rm Ni_{62}Nb_{38}$~\cite{zhang2016} and $\rm Zr_{60}Cu_{29}Al_{11}$~\cite{cheng2009}, and a pairwise Lennard-Jones potential for $\rm Ni_{62}Nb_{38}$. (The energetic and size parameters and details of the simulation methods for the LJ potential are given in the Supplemental Material (SM) ~\cite{SM}).)
Each metallic glass sample contains $N=6000$ atoms in a 2$\times$2$\times$3 parallelepiped with periodic boundaries in the $x$-, $y$-, and $z$-directions. The samples are produced by quenching equilibrium liquid states at $T=2000$~K (above the glass transition) to $T_0=2$~K using a range of cooling rates from $R=10^{10}$ to $10^{13}$ K/s. 
The temperature of the system is defined as $T=2 {\cal K}/(3N k_{\rm B})$, where ${\cal K}$ is the kinetic energy and $k_{\rm B}$ is the Boltzmann constant.
After quenching to low temperature, the periodic boundary conditions in the $x$- and $y$-directions are changed to open boundary conditions, and the system is relaxed using the Nos\'e-Hoover thermostat and barostat (in the $z$-direction) to achieve zero pressure at $T_0$.

We apply uniaxial compression in the $z$-direction. We define the engineering strain and stress as  
\begin{eqnarray}
  \varepsilon_{\rm eng} & =& (L_{z0}-L_z)/L_{z0},\\
  \sigma_{\rm eng} & =& F_z/A_0,
\end{eqnarray}
where $L_{z0}$ is the undeformed length of the sample in the $z$-direction, $A_0$ is the undeformed cross-sectional area in the $x$-$y$ plane, and $F_z$ is the total force in the $z$-direction that opposes the compression. To investigate heating during the applied deformation, we compress the system at constant strain rate ${\dot \varepsilon}_{\rm eng}$ and apply a viscous damping force proportional to the velocity ${\bm v}_i$ of atom $i$, ${\bm F}_i=-b{\bm v}_i$ to dissipate the energy input from compression. We vary the damping coefficient from $b=0$ to $10^{-3}$ $\rm eV\cdot ps/\text{\r{A}}^2$.  The strains are applied affinely to the system, such that after each compressive strain increment, $\Delta\varepsilon_{\rm eng} = {\dot \varepsilon}_{\rm eng} \Delta t$, the atomic $z$-positions are scaled as $z(1-\Delta \varepsilon_{\rm eng})$, where $\Delta t$ is the time step.

\begin{figure}[tbp]
\centering
\includegraphics[width=0.45\textwidth]{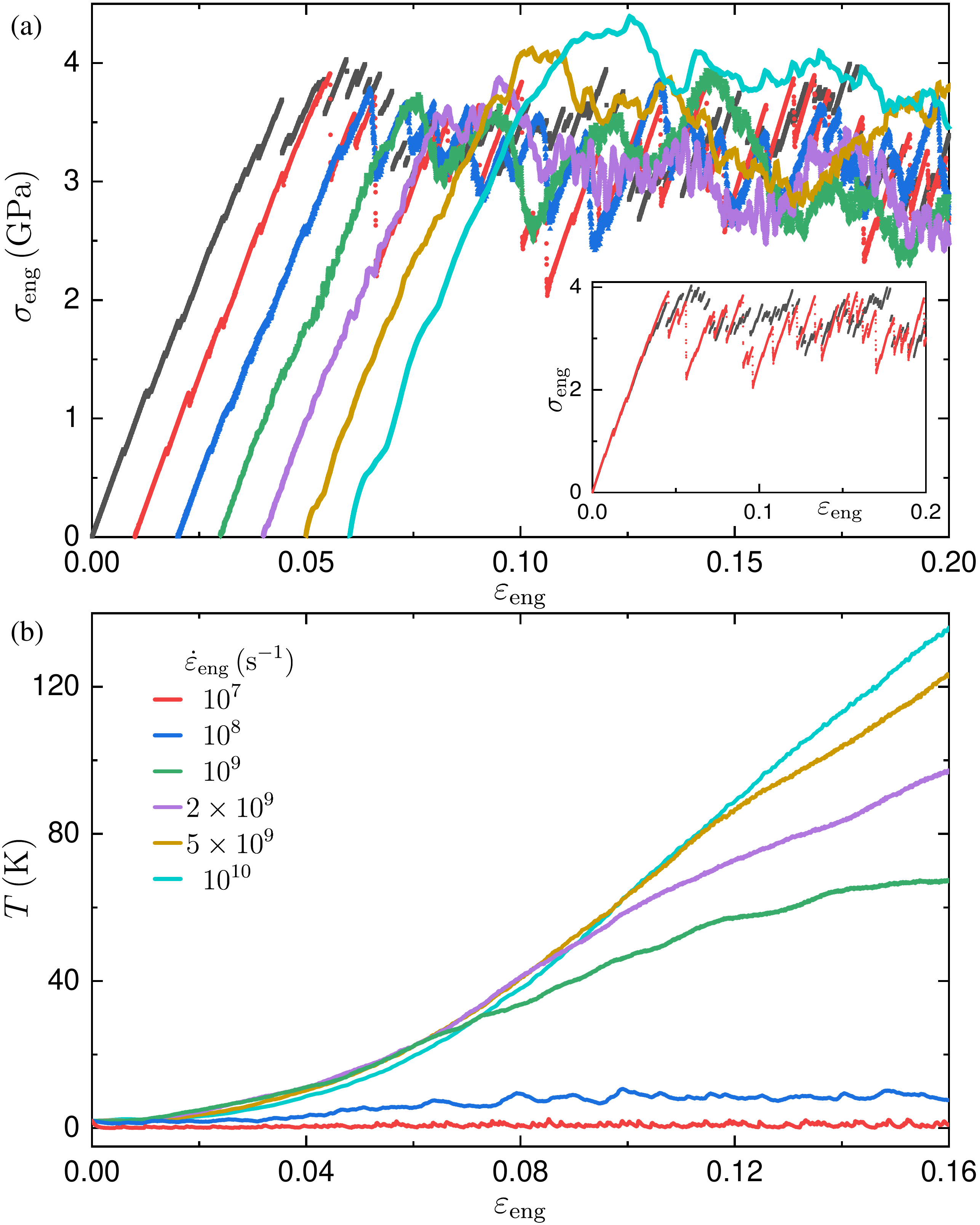}
\caption{(a) Engineering stress $\sigma_{\rm eng}$ versus strain $\varepsilon_{\rm eng}$ from EAM simulations of uniaxial compression of a $\rm Ni_{62}Nb_{38}$ sample obtained using cooling rate $R = 10^{12}$~K/s.  The black curve corresponds to athermal quasistatic compression and the horizontally shifted curves correspond to compression  with damping coefficient $b=10^{-4}$ $\rm eV\cdot ps/\text{\r{A}}^2$ and strain rates ${\dot \varepsilon}_{\rm eng}=10^7$ (red), $10^8$ (blue), $10^9$ (green), $2\times10^9$ (magenta), $5\times10^9$ (orange), and $10^{10}$ (cyan) $\rm s^{-1}$. The inset shows the same data for $\sigma_{\rm eng}$ versus $\varepsilon_{\rm eng}$ in the main panel without horizontally shifting the curve with ${\dot \varepsilon}_{\rm eng}=10^7$ $\rm s^{-1}$. (b) Average temperature $T$ versus strain $\varepsilon_{\rm eng}$ from EAM simulations of twenty $\rm Ni_{62}Nb_{38}$ samples prepared using $R = 10^{12}$~K/s for each strain rate. The damping coefficient and strain-rate color codes are the same as those in (a).}
\label{figStrainrate}
\end{figure}

{\em Results.}---In this section, we present the results from EAM simulations of $\rm Ni_{62}Nb_{38}$ undergoing uniaxial compression. The results for the EAM simulations of $\rm Zr_{60}Cu_{29}Al_{11}$ and LJ simulations of $\rm Ni_{62}Nb_{38}$ are qualitatively the same and are presented in the SM \cite{SM}. In Fig.~\ref{figStrainrate} (a), we show the stress versus strain relation for uniaxial  compression at a fixed damping coefficient 
$b=10^{-4}$~$\rm eV\cdot ps/\text{\r{A}}^2$ and several strain rates. At low strain rates, the stress-strain curves possess quasi-linear elastic segments, punctuated by discontinuous drops in the stress. In particular, the stress-strain curves for ${\dot \varepsilon}_{\rm eng} \lesssim 10^{8}$ $\rm s^{-1}$ agree with those obtained using athermal quasistatic compression. Serrations in the stress-strain relations have also been found in experimental studies of uniaxial compression and nanoindentation of BMGs at low strain rates~\cite{ramachandramoorthy2021,yu2021}. The stress-strain curves become more continuous and the maximum stress is non-monotonic with increasing strain rate. For ${\dot \varepsilon}_{\rm eng} =10^{8}$~ $\rm s^{-1}$, the maximum stress value decreases below that achieved in the low strain-rate limit. As the strain rate increases further, i.e. for ${\dot \varepsilon}_{\rm eng} \gtrsim 5 \times 10^{9}$~$\rm s^{-1}$, the maximum stress is larger than that in the low strain-rate limit. 
In Fig.~\ref{figStrainrate} (b), we show the variation of the internal temperature with strain at fixed damping coefficient and several strain rates. For most strain rates, the temperature increases from $T_0$ at zero strain to a value $T_f$ at larger strains $\varepsilon_f$, where both $\varepsilon_f$ and $T_f$ increase with strain rate. For ${\dot \varepsilon}_{\rm eng} \gtrsim 2\times 10^{9}$~ $\rm s^{-1}$, we do not show the strain regime where $T \rightarrow T_f$.

\begin{figure}[tbp]
\centering
\includegraphics[width=0.45\textwidth]{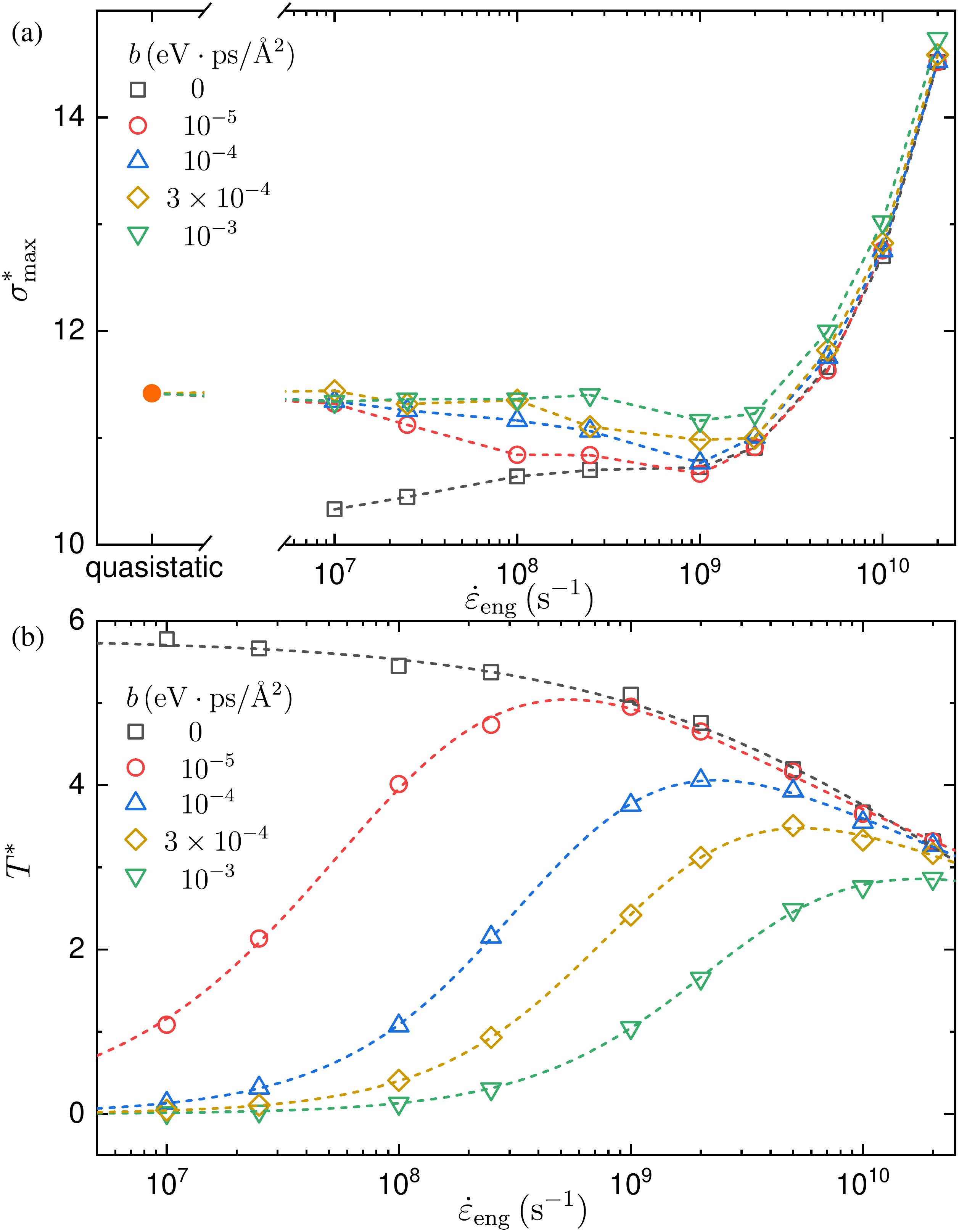}
\caption{(a) Compressive strength $\sigma^*_{\rm max}$ (scaled by the cooling rate used to prepare the samples) plotted versus strain rate ${\dot \varepsilon}_{\rm eng}$ for $\rm Ni_{62}Nb_{38}$ (averaged over $80$ samples) at several damping coefficients $b$. The data for athermal quasistatic compression is shown as a filled circle. (b) The mean temperature ${T}^*$ (obtained in the strain interval where the maximum stress occurs and scaled by the cooling rate used to prepare the samples) versus ${\dot \varepsilon}_{\rm eng}$ for the systems in (a). Best fits of $T^*({\dot \varepsilon}_{\rm eng})$ to  Eq.~(\ref{eqTvsSrate}) for different $b$ values are shown as dashed lines. }
\label{figMaxStressAvgT}
\end{figure}

As shown above, the stress-strain relations for metallic glasses depend on the damping coefficient and strain rate, as well as the cooling rate.  The results for the compressive strength as a function of the strain rate for EAM simulations of $\rm Ni_{62}Nb_{38}$ prepared using different cooling rates and compressed using different damping coefficients are shown in Fig. S5 (a) in the SM \cite{SM}.
We find that the compressive strength $\sigma_{\rm max}$ has a self-similar form for different cooling rates. We therefore plot the cooling-rate scaled compressive strength $\sigma^*_{\rm max}=\langle\sigma_{\rm max}/(\frac{R}{R_c})^{-\gamma}\rangle_R$, where $\gamma \sim 0.038$ and $R_c$ is a reference cooling rate, versus strain rate in Fig.~\ref{figMaxStressAvgT} (a). 
$\langle...\rangle_R$ indicates the average of $\sigma_{\rm max}/(\frac{R}{R_c})^{-\gamma}$ over the four cooling rates, $R=10^{10}$, $10^{11}$ $10^{12}$, and $10^{13}$ K/s.
We find several important features for $\sigma^*_{\rm max}({\dot \varepsilon}_{\rm eng})$ . First, in the low strain-rate limit (for the damping coefficients considered), $\sigma^*_{\rm max}$ approaches the value obtained for athermal quasistatic compression. Second, the scaled compressive strength increases monotonically with strain rate when the damping coefficient $b=0$, whereas $\sigma^*_{\rm max}$ is nonmonotonic in ${\dot \varepsilon}_{\rm eng}$ when $b>0$~\cite{ramachandramoorthy2021}.
For nonzero damping coefficients, we find that the scaled compressive strength first decreases with increasing strain rate and then increases rapidly for ${\dot \varepsilon}_{\rm eng} \gtrsim 10^9$ $\rm s^{-1}$. The magnitude of the nonmonotonic behavior (i.e. the difference between the value of $\sigma^*_{\rm max}$ as ${\dot \varepsilon}_{\rm eng} \rightarrow 0$ and the minimal value of $\sigma^*_{\rm max}({\dot \varepsilon}_{\rm eng})$) increases with decreasing $b$.

To understand the nonmonotonic behavior of  $\sigma^*_{\rm max}$ versus ${\dot \varepsilon}_{\rm eng}$, we characterize the temperature of the system during compression. In Fig.~\ref{figMaxStressAvgT} (b), we plot the temperature averaged over strains in the range $0 < \varepsilon_{\rm eng} < \varepsilon_b$ and scaled by the cooling rate, $T^*= \langle T/(\frac{R}{R_c})^{\lambda}\rangle_R$, where $\lambda \approx 0.049$~\cite{SM}. (For $\varepsilon_b \sim 0.084$, $\sigma^*_{\rm max}$ occurs in this strain interval for all systems considered.)  The temperature versus strain rate can be captured by the expression 
\begin{equation}
   {T}^*=\frac{c ({\dot \varepsilon/k})^\alpha}{1+({\dot \varepsilon}/k)^\beta},
\label{eqTvsSrate}
\end{equation}
where the coefficients $k$ and $c$ have units of ${\rm s}^{-1}$ and ${\rm K}$, respectively, and the power-law exponents satisfy $\beta >\alpha$ and depend on $b$. Note that $\alpha=0$ for systems with $b=0$. 
The scaled average temperature ${ T}^*$ decreases (and the scaled maximum stress increases) monotonically with increasing strain rate for systems with zero damping coefficient, since the averaging strain interval is fixed and the system has less time to heat up as the strain rate increases. In contrast, ${ T}^*$ versus ${\dot \varepsilon}_{\rm eng}$ possesses a maximum, whose height decreases and position in ${\dot \varepsilon}_{\rm eng}$ increases as the damping coefficient increases. 
At low strain rates (before the peak in ${ T}^*({\dot \varepsilon}_{\rm eng})$), the damping can effectively remove heat from the system during compression, and thus the temperature at low strain rates decreases with increasing damping coefficient. As the strain rate is increased, damping is less effective at removing heat and the temperature increases with strain rate.  For large strain rates after the peak in ${ T}^*({\dot \varepsilon}_{\rm eng})$, the strain rate is so fast that sufficient time has not elapsed to allow the increased potential energy from compression to be converted into kinetic energy, and thus the temperature decreases. Note that we are not interested in the behavior of ${ T}^*({\dot \varepsilon}_{\rm eng})$ for extremely large strain large rates ${\dot \varepsilon}_{\rm eng} \gtrsim 10^{10}$~${\rm s}^{-1}$, since this regime of rapidly increasing $\sigma^*_{\rm max}$ is difficult to achieve in experiments on BMGs.  

The results in Fig.~\ref{figMaxStressAvgT} suggest that the nonmonotonic behavior of $\sigma_{\rm max}$ versus ${\dot \varepsilon}_{\rm eng}$ is caused by the nonmonotonic dependence of temperature on strain rate. Therefore, we plot the scaled compressive strength $\sigma^*_{\rm max}$ versus the scaled mean temperature ${ T}^*$ for EAM simulations of $\rm Ni_{62}Nb_{38}$ in Fig.~\ref{figExpSim} (b) (as highlighted in Fig. S6 in SM \cite{SM}). 
In the limit of small damping coefficient ($b\rightarrow0$), $\sigma^*_{\rm max}$ versus ${ T}^*$ follows a chevron-shaped curve, which consists of two regimes. For the low strain-rate regime, $\sigma^*_{\rm max}$ decreases linearly with increasing $T^*$ and in the high-strain rate regime the compressive strength increases as a power-law with decreasing temperature, $\sigma^*_{\rm max} \sim (T^*)^{-\beta}$, where $\beta \approx 4.8$.  For higher damping coefficients, the temperature dependence of $\sigma^*_{\rm max}$ deviates from the linearly decreasing asymptotic behavior at a $T^*$ that depends on $b$. At high strain rates, $\sigma^*_{\rm max}$ versus $T^*$ converges to the power-law asymptotic behavior for all $b$.  

We also performed EAM simulations of uniaxial compression of $\rm Zr_{60}Cu_{29}Al_{11}$ and found similar behavior for $\sigma_{\rm max}$ and $T$ versus ${\dot \varepsilon}_{\rm eng}$. (See Figs. S7 and S8 in SM~\cite{SM}.) In particular, in the small-damping limit, we find a similar chevron-shaped curve for $\sigma^*_{\max}$ versus $T^*$. Again, $\sigma^*_{\max}$ decreases linearly with increasing $T^*$ in the low strain-rate asymptotic regime and grows as a power-law with decreasing $T^*$ with exponent $\beta \approx 5$ in the high strain-rate asymptotic regime. 

Finally, we performed uniaxial compression studies of $\rm Ni_{62}Nb_{38}$ using a Lennard-Jones potential. (See Figs. S2-S4~\cite{SM} in SM.). Again, we find similar behavior for $\sigma_{\rm max}$ and $T$ versus ${\dot \varepsilon}_{\rm eng}$. However, the magnitude of the slope of the linear asymptotic regime is half of that from the EAM simulations of $\rm Ni_{62}Nb_{38}$ and the power-law exponent of the high strain-rate asymptotic regime is $\beta \approx 9.6$, which is roughly twice the value from the EAM simulations of $\rm Ni_{62}Nb_{38}$. One possible explanation for the difference in the slopes and power-law exponents of the asymptotic behavior is the difference in the specific heat obtained for the LJ and EAM potentials of $\rm Ni_{62}Nb_{38}$. 

We now compare the experimental data on uniaxial compression of BMGs~\cite{
zhang2007,ma2009,chen2015,amit2022,mukai2002,xue2008,zheng2011,li2017,wang2014,lu2003,liu2010,ramachandramoorthy2021} in Fig.~\ref{figExpSim} (a) to the EAM simulation data for $\rm Ni_{62}Nb_{38}$ in Fig.~\ref{figMaxStressAvgT}.  (We obtain similar results when we compare the experimental data in Fig.~\ref{figExpSim} (a) to the EAM simulation data for $\rm Zr_{60}Cu_{29}Al_{11}$.)
We first scale the maximum stress and strain rate for each experimental data set, 
$\sigma_{\rm max}^{'}=k_{\sigma,e}\sigma_{\rm max}+c_{\sigma,e}$
 and ${{\dot \varepsilon}^{'}_{\rm eng}}=  ({\dot\varepsilon}_{\rm eng}/k_{\varepsilon,e})^{\eta_{\varepsilon,e}}$, choosing the constants $k_{\sigma,e}$, $c_{\sigma,e}$, and $k_{\varepsilon,e}$, exponent ${\eta_{\varepsilon,e}}$, which are provided in SM \cite{SM}, and damping coefficient $b$ that give the best fit to the EAM simulation data for $\rm Ni_{62}Nb_{38}$.  
 After determining the best-fit damping coefficient, and the maximum stress and strain rate scaling, we can identify the effective $T^*$ versus ${\dot \varepsilon}_{\rm eng}$ relation (see Fig.~\ref{figMaxStressAvgT} (b)) for each experimental data set. We can then eliminate ${\dot \varepsilon}_{\rm eng}$ from the expressions for $\sigma^{*}_{\rm max}({\dot \varepsilon}_{\rm eng})$ and $T^{*}({\dot \varepsilon}_{\rm eng})$ to obtain $\sigma^*_{\rm max}$ versus $T^*$ for the experimental data. The scaled compressive strength $\sigma^*_{\rm max}$ versus temperature $T^*$ curves collapse onto the EAM simulation results for $\rm Ni_{62}Nb_{38}$ for different $b$ values as shown in Fig.~\ref{figExpSim} (b).

{\em Discussion.}---We have shown that the compressive strength versus temperature relation is universal for a wide range of metallic glasses undergoing uniaxial compression at finite strain rates, and that this chevron-shaped universal relation controls their mechanical response. The experimental data for $12$ different BMGs, EAM simulations of two BMGs $\rm Ni_{62}Nb_{38}$ and $\rm Zr_{60}Cu_{29}Al_{11}$, and LJ simulations of $\rm Ni_{62}Nb_{38}$ can all be scaled onto similar master curves for $\sigma^*_{\rm max}$ versus $T^*$, which determine the compressive strength versus strain rate.  At low strain rates, increasing the strain rate leads to increasing temperature and {\it decreases} in $\sigma^*_{\rm max}$, whereas at high strain rates, increases in strain rate lead to decreasing temperature and {\it increases} in  $\sigma^*_{\rm max}$.  This non-monotonic behavior of $\sigma^*_{\rm max}$ versus temperature causes the nonmonotonic behavior of $\sigma^*_{\rm max}$ versus strain rate. Variations in the internal dissipation change the characteristic strain rate at which the nonmonotonic behavior occurs. These results are general for a wide range of metallic glasses with different atomic interaction potentials, damping coefficients, and chemical compositions. 

In this work, we focused on uniaxial compression of metallic glasses at non-zero strain rates. However, we believe that our results will also hold for other deformations, such as simple and pure shear and indentation, applied to metallic glasses at finite rates.  Further, our work emphasizes that internal heating and dissipation mechanisms control the strain-rate-dependent mechanical response. Thus, future computational studies can investigate the non-affine collective motions of atoms, or shear transformation zones, that give rise to local heating and dissipation. For example, in recent work~\cite{jin2021} we developed an exact method to identify and track local deformations during stress drops that result from athermal, quasistatic simple shear applied to model glasses. In future studies, we will generalize the methods for identifying shear transformation zones~\cite{sopu17,peng2011,hassani2019,im2021} in metallic glasses deformed at finite rates and temperature. 

\begin{acknowledgments}
We acknowledge support from NSF Grant Nos. CMMI-1901959 (W.J., A.D., U.D.S., and C.S.O.) and CBET-2002797 (M.D.S.).
This work was also supported by the High Performance Computing facilities operated by Yale’s Center for Research Computing.
\end{acknowledgments}

\end{document}